# Operation of a LAr-TPC equipped with a multilayer LEM charge readout

B. Baibussinov[1], S. Centro[1], C. Farnese[1], A. Fava[1a], D. Gibin[1], A. Guglielmi[1], G. Meng[1], F. Pietropaolo[1,2], F. Varanini[1], S. Ventura[1], K. Zatrimaylov[1, 3b]

[1] *INFN Sezione di Padova and Dip. di Fisica e Astronomia Università di Padova, Padova, Italy*
[2] *CERN, European Laboratory for Particle Physics, Geneva, Switzerland*
[3] *Novosibirsk State University and Budker Institute of Nuclear Physics, Novosibirsk, Russia.*

*E-mail*: `Francesco.Pietropaolo@cern.ch`

ABSTRACT: A novel detector for the ionization signal in a single phase LAr-TPC, based on the adoption of a multilayer Large Electron Multiplier (LEM) replacing the traditional anodic wire arrays, has been experimented in the ICARINO test facility at the INFN Laboratories in Legnaro. Cosmic muon tracks were detected allowing the measurement of energy deposition and a first determination of the signal to noise ratio. The analysis of the recorded events demonstrated the 3D reconstruction capability of ionizing events in this device in liquid Argon, collecting a fraction of about 90% of the ionization signal with signal to noise ratio similar to that measured with more traditional wire chambers.

---

[a] Now at Fermilab
[b] Now at Scuola Normale Superiore, Pisa

**Contents**

### 1. Introduction

Liquid Argon Time Projection Chamber (LAr-TPC) detection technique [1] has been continuously evolving in the last decades and is presently worldwide acknowledged as well suited for the study of particle physics of rare events, especially in the neutrino oscillation field. LAr-TPCs provide high granularity imaging by collecting with mm pitch anodic wire chambers the ionization electrons, produced by the interaction of charged particles with the target medium and drifted by a uniform electric field over long drift paths. The state of the art is the ICARUS-T600 detector, the biggest LAr-TPC ever built with 760 t LAr mass, installed underground at the INFN Gran Sasso Laboratories and exposed to the CNGS neutrino beam from CERN in the 2010-2012 period [2] [3]. Nevertheless, intense R&D activities are ongoing to further improve the performance especially in terms of signal to noise ratio (S/N), presently assessed at 10:1 level with warm electronics [3] and mainly limited by the impossibility to achieve a multiplication of the ionization electrons in the liquid phase.

A modified version of the traditional LAr-TPC chamber has been developed in the contest of double phase LAr-TPC detectors [4] by substituting the usual wire arrays with multiple Large Electron Multiplier (LEM) planes as derived from the Gas Electron Multiplier (GEM) detectors [5].

As an alternative a single phase multilayer LEM prototype has been realized and installed in the ICARINO LAr-TPC test-facility INFN-LNL Laboratories in Legnaro [6] replacing the standard anodic wires arrays. A cosmic rays data taking campaign to qualify the performance of this single phase LEM LAr-TPC to detect the signals from ionization electrons drifting to the TPC anode has been done. The test has been dedicated to assess the viability to detect the drifting charge generated in ionizing events in single-phase LAr-TPC with a multilayer LEM, i.e. the achievement of a good S/N while maintaining the millimeter detector granularity of wire chambers, preserving at the same time the possibility of non-destructive readout of drifting electrons over multiple parallel planes. The last feature, together with the adoption of different orientation of the LEM electrode strips, is required for an accurate 3D reconstruction of ionizing events in the LAr [7].

In the present paper the results of this first attempt to operate a multilayer LEM LAr-TPC to detect cosmic muons are reported.

## 2. Description of the experimental set up

The ICARINO set-up includes a fully functional small LAr-TPC operated in INFN-LNL for test purposes (Figure 1). The liquid argon (LAr) is contained in a stainless steel cylindrical vessel closed by an ultra-high vacuum flange hosting the feed-through for vacuum, LAr filling and re-circulation, high voltage cables and readout electronics. The whole detector vessel is contained in an open-air stainless steel dewar, which is initially filled with commercial LAr acting as cryogenic bath for the ultra-pure LAr injected in the detector vessel. Details of the cryogenic system and the procedures for LAr filling and recirculation can be found elsewhere [6].

The standard TPC chamber active volume corresponds to 38 kg of active LAr mass, delimited laterally by 4 29.4x29.4 $cm^2$ FR4 boards supporting the field-shaping electrodes (30 strips of gold-plated copper) and by two vertical electrode planes, 32.6x32.6 $cm^2$, acting as cathode and anode (Figure 2 left). As a default the drift field is fixed to the standard value of

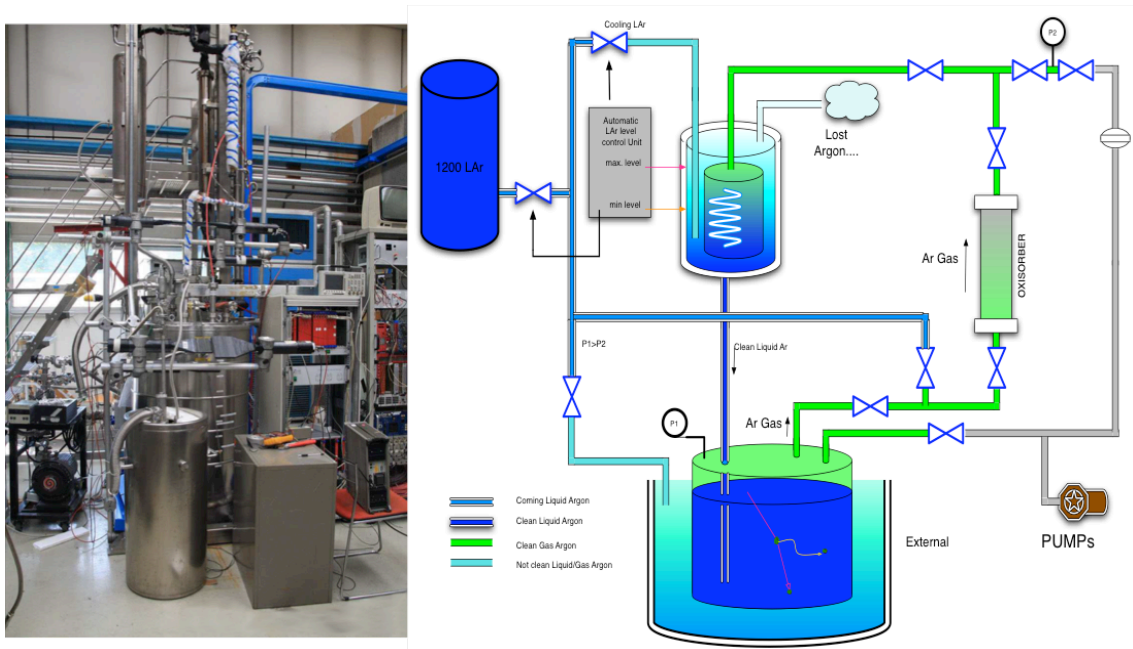

**Figure 1 The ICARINO set-up in INFN-LNL (left) and the scheme of the cryogenic plant (right).**

500 V/cm by applying an appropriate HV to the cathode plane. For the present test the anodic wire chamber has been replaced with the multilayer LEM detector described later (Figure 2 right) mechanically fitting the TPC chamber structure without any modification.

The system is equipped with a sub-set of the standard ICARUS DAQ system [2], designed to provide continuous digitization and waveform recording with a 32-channel modularity. Each signal is passed through a decoupling capacitor to the CAEN V791 analogue board, which hosts the front-end amplifiers (1000 electrons per ADC count gain) and performs 16:1 multiplexing and 10-bit ADC digitization at 40 MHz, so that the time sample per each channel corresponds to 400 ns. Differently from the ICARUS-T600 configuration, both Induction and Collection planes have been read with the CAEN V791Q boards in current-like mode with a 3 μs time constant signal integration to preserve the bipolar and unipolar shape respectively.

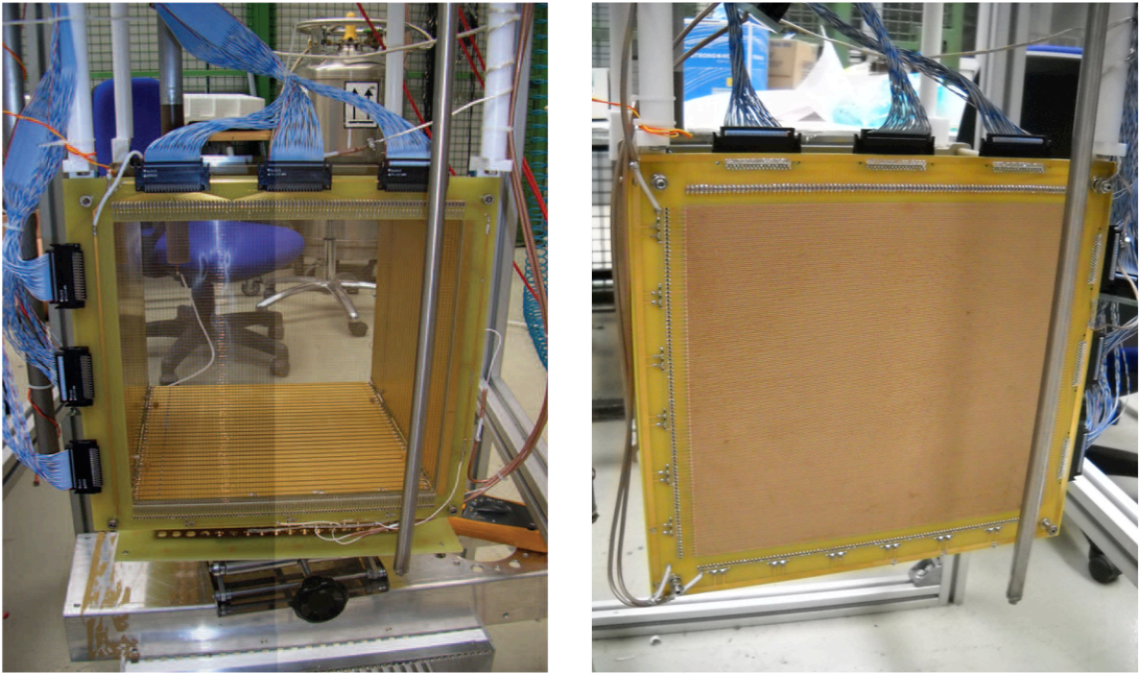

**Figure 2 The TPC chamber equipped with standard anodic vertical wire planes (left) and with the corresponding new multilayer LEM plane (right).**

Data are continuously recorded on the CAEN V789 digital boards in multi-event circular memory buffers, upon application of a standard lossless compression algorithm that allows to reduce the size by a factor 4 [8]. The buffer size has been set to 1024 t-samples to ensure recording of a full drift time (about 500 t-samples at the nominal electric field) in addition to enough overhead, needed f.i. for controlling the noise conditions. When a trigger signal occurs, the active buffer is frozen and the stored data are readout by the DAQ.

The experimental setup is complemented by a telescope composed by set of 4 external plastic scintillators for triggering purposes, selecting crossing cosmic muons in 30°÷50° zenith angle, so that most of the collection view channels and a large fraction of the induction view ones were interested in the event.

## 2.1 The multilayer LEM anodic plane

Characteristics and performance of standard GEM on double-face Printed Circuit Board (PCB), depending on the size and density of holes and thickness of PCB, are widely described in literature (see f.i. [5]).

A new cryogenic multilayer LEM detector has been realized and installed in the present ICARINO set-up to detect the ionization charge directly in high purity liquid argon. The device is composed by a standard printed-circuit board, 3.2 mm thick, with sub-millimeter holes (0.5 mm diameter and 1 mm spacing) on ~23% of the surface, chemically etched at their rims using automatic micromachining (Figure 3 left). The PCB board has 3 electrical layers (Figure 3 right), the first one acting as a screening grid, while the other two are used to detect the drifting electrons. The two read-out planes, spaced by 1.6 mm, are made of an array of 96 copper strips 3 mm wide. The first, with the strips vertically oriented, works in Induction mode, while the second collects the drifting electrons on its horizontal electrodes (Figure 4). Each strip contributes ~18 pF capacitance to the preamplifier input, to be compared to the 47 pF of the cable itself.

The grid plane, 1.6 mm in front of the Induction plane, was electrically biased at -220 V to ensure the homogeneity of the inner drift field, acting as an electromagnetic shield to improve the Induction signal sharpness and reduce the noise from the HV cathode biasing. Finally, another thin stainless steel grid was inserted behind the Collection plane and grounded to confine the electric field in close proximity of the anodic electrodes. A -15.7 kV voltage is applied to the cathode, corresponding to a uniform electric field of 500 V/cm in the drift volume.

With an appropriate biasing the electric field between successive measuring planes is pinched into the detector holes to produce the natural channeling of the drift charge through the holes themselves, ensuring the full transparency of multiple planes. The hole diameter being much smaller than the strip pitch size ensures that the signals induced by the charge drifting through each hole is almost completely restricted to a single strip, minimizing the image blurring due to cross talk between adjacent strips as well as the crossing time jitter. Moreover the geometry of the detector ensures a more complete screening by each individual plane.

As a first approximation the transparency condition is reached when the ratio between the electric fields in the drift region and in the holes is equal to the fraction of pierced surface. This geometrical condition corresponds in the device under test to ~350 V gap between adjacent planes. The passing charge will induce a bipolar signal in the middle Induction plane and then be integrally collected in the Collection strips. An example of the waveforms recorded on an Induction and Collection strip is shown in Figure 6.

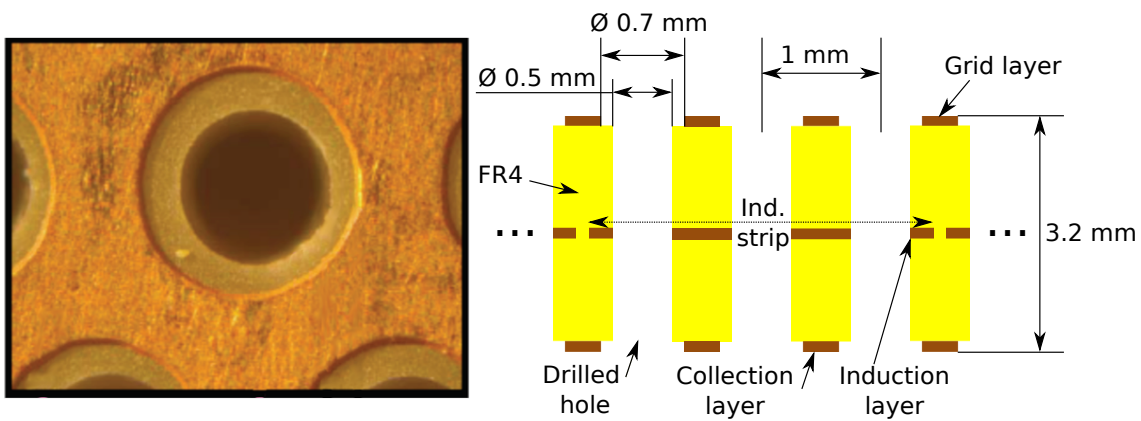

**Figure 3 Left: zoomed photo of one hole (0.5 mm) seen from the grid side and security ring (0.7 mm) of the LEM PCB board. Right: sketch of LEM cross section along a Collection strip, in correspondence of the hole centers (Induction strips are orthogonal to the page).**

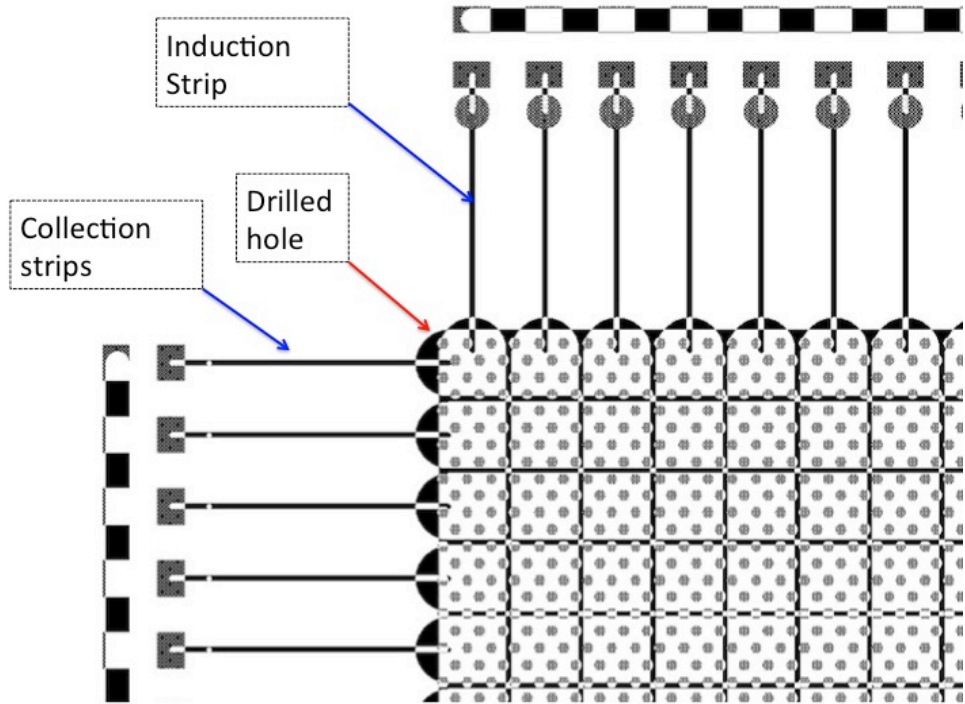

**Figure 4 Detail of the upper corner of the LEM board, showing the hole pattern, the horizontal (Collection) strips and, in transparency, the vertical (Induction) readout strips and the soldering pads.**

## 3. Test with cosmic rays

After the initial commissioning phase dedicated to reach an adequate LAr purity, the LAr LEM-TPC was operated in steady conditions for a week, recording the passing cosmic rays. The "electron lifetime" [5] [9] which characterizes the exponential reduction of drift electrons from the cathode to the anode turned out to be always exceeding few milliseconds, guaranteeing a negligible attenuation of the electron signal in the measured drift range.

The adopted trigger was selecting particles at ~40° zenith angle, corresponding to an effective average pitch of 4 mm per Collection strip. Almost 10000 triggered events were collected. The reference electric biasing was set to ΔV=350 V between each of the three consecutive LEM layers; different polarization conditions were also tested (Table 1).

| $V_{Grid}$ (V) | $V_{Ind}$ (V) | $V_{Coll}$ (V) | MPV signal ($e^-$/mm) |
|---|---|---|---|
| -220 | 80 | 380 | 4550 |
| -220 | 130 | 480 | 4800 |
| -220 | 180 | 580 | 5000 |

**Table 1 The electric polarization bias applied to the LEM layers during the test. The corresponding most probable value (MPV) of the electron signal in Collection produced by m.i.p. muons is also reported in the last column.**

## 3.1 Cosmic ray data reconstruction and analysis

At this stage the tests were mostly concentrated on the detection and optimization of the signals in Collection in view of the calorimetric measurement of events. In addition the signals in Induction involved in the stereoscopic reconstruction of events, were also studied. A clean sample of passing cosmic muons was selected reconstructing 3D tracks based on the time matching of hits in the two available projections. A typical event collected with the present setup is displayed in Figure 5 for the Induction (right) and Collection (left) planes, demonstrating imaging capabilities comparable to the more standard wire chamber LAr-TPCs. The corresponding typical signals on a Collection and an Induction strip are shown in FiFigure 6. The electronic noise was measured in collection view to have an rms of ~1.7 ADC counts. corresponding to a signal to noise ratio of about 9 for passing cosmic muons.

The signals on each strip have been identified according to the standard ICARUS method [2]. An automatic procedure has been specifically developed, optimizing the algorithms already exploited in the analysis of ICARUS-T600 [8]. The geometrical reconstruction has been obtained by building the cluster of reconstructed hits separately in the two orthogonal projections and associating them on the basis of their drift times. The automatic procedure has been visually checked on a subsample of events to correctly reconstruct the cosmic muons tracks crossing the LAr-TPC volume. The amount of charge signal on each Collection strip depends on the "pitch", i.e. the length of the track segment seen by each strip, 4 mm on average in the collected sample. The selected tracks have been required to extend over at least 30 collection strips and 30 t-samples in the drift direction. The integrated signals, normalized to 1 mm track segment, are distributed according to a Landau convoluted with a Gaussian to account for the electronic noise and the spread of muon momentum (Figure 7). The calibration factor of the ICARUS electronic transfer function is assumed to be 0.038 fC/(ADC x μs) as measured for the T600 LAr-TPC [10].

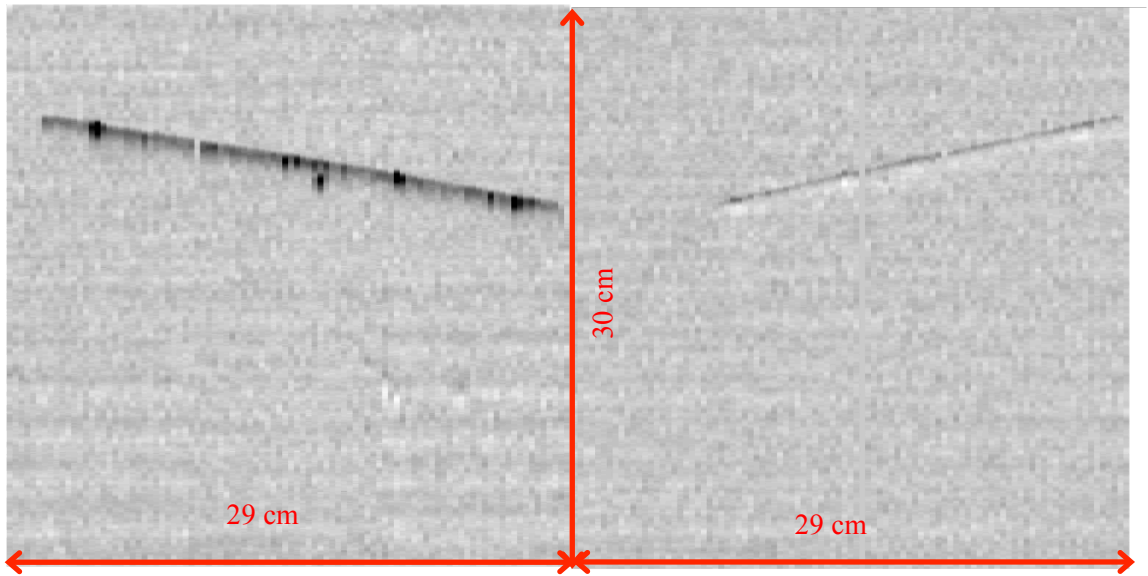

**Figure 5 A typical event collected in the TPC in collection (left) and induction (right) projections. The muon is entering with an angle of 39° from the upper side of the chamber.**

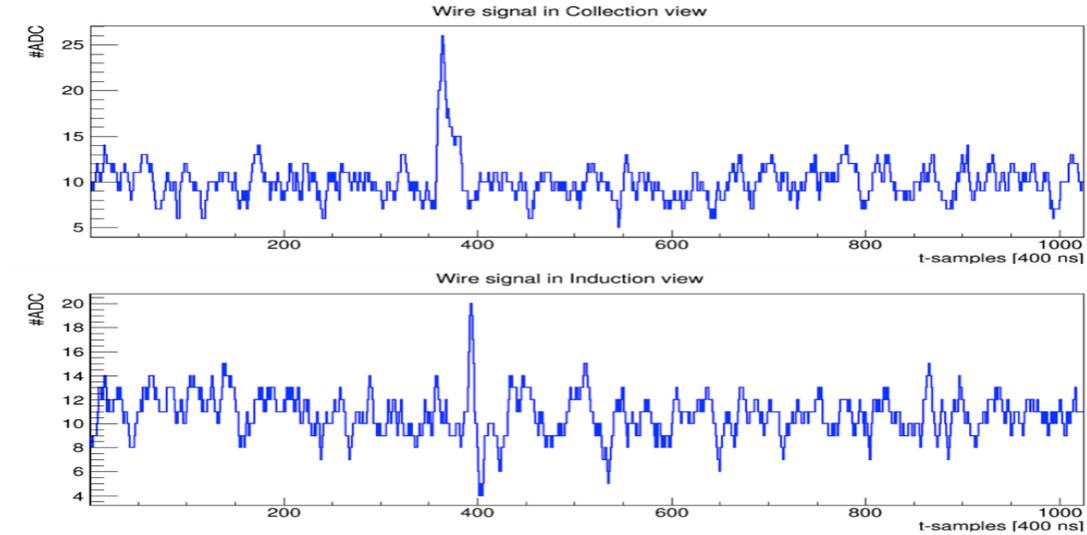

**Figure 6 Typical signals on a Collection (top) and Induction (bottom) LEM strips.**

Measurements demonstrated that the reference 350 V electric bias was not sufficient to reach the full transparency condition and that instead the signal was not yet saturated and reaching a value of ~5000 e$^-$/mm at $\Delta V=400$ V electric bias (Table 1 and Figure 7). As a reference, past measurements [6] with the same LAr-TPC, equipped with a standard wire chamber, performed with cosmic muons crossing at a similar $\theta = 45°$ angle, i.e. with a pitch of 4.2 mm, provided a most probable fit value of 240 (ADC * 0.4 μs), corresponding, after the application of the above mentioned calibration factor, to ~5500 e$^-$/mm.

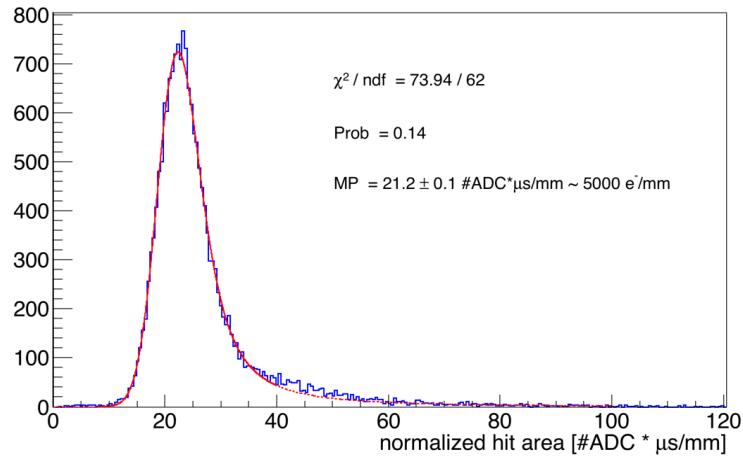

**Figure 7 Distribution of the signals on collection strips normalized to 1 mm pitch ($\Delta V=400$ V) (blue histogram) with the Landau Gaussian fit (red curve) for the selected cosmic muons. The fit has been restricted to signals below ~2 MPV, to exclude the possible contribution due to overlapped delta rays.**

Therefore the LEM board with $\Delta V=400$ V allowed to collect a fraction of ~90% of the ionization signal measured with standard wire chamber. As a next step additional studies in view of the LEM biasing optimization are required.

## 4. Conclusions

The successful operations of a LAr-TPC equipped with a multilayer LEM anodic plane proved its feasibility to detect and measure charge signals in the single phase LAr-TPC. Tests with passing cosmic rays demonstrated the possibility to obtain a 3D reconstruction by non-destructive sensing of the drifting electrons over successive planes.

The measured signals in Collection view appeared close to the ones of traditional wire chambers, even if a full transparency condition wasn't reached for electric biases close to the value expected from a naïve geometrical approximation. A ~9 signal to noise ratio was also measured in Collection, similar to the corresponding value in ICARUS at the LNGS.

In the present experimental test the major emphasis was put on the study of the signal detection in the Collection plane. Signals from the intermediate Induction plane were only used for the 3-D geometrical reconstruction of crossing tracks.

A successive optimization phase of the LEM design could be foreseen. Moreover fast readout electronics could help to better qualify the signals in the intermediate Induction planes. The practical adoption of multilayer LEM in large LAr-TPCs still requires detailed R&D to define how to pave the anodic planes with the necessary mechanical accuracy and the possible wiring schemes.